# E-Commerce Product Recommendation System based on ML Algorithms.

Md. Zahurul Haque, Department of CSE, Manarat International University, Dhaka, Bangladesh.
jahurulhaque@manaratr.ac.bd

*Abstract: Algorithms are used in e-commerce product recommendation systems. These systems just recently began utilizing machine learning algorithms due to the development and growth of the artificial intelligence research community. This project aspires to transform how ecommerce platforms communicate with their users. We have created a model that can customize product recommendations and offers for each unique customer using cutting-edge machine learning techniques, we used PCA to reduce features and four machine learning algorithms like Gaussian Naive Bayes (GNB), Random Forest (RF), Logistic Regression (LR), Decision Tree (DT), the Random Forest algorithms achieve the highest accuracy of 99.6% with a 96.99 r square score, 1.92% MSE score, and 0.087 MAE score. The outcome is advantageous for both the client and the business. In this research, we will examine the model's development and training in detail and show how well it performs using actual data. Learning from machines can change of ecommerce world.*

## I. INTRODUCTION

**E-**commerce, also known as electronic commerce or online e-commerce, refers to the purchase and sale of products or services through the internet, as well as the transmission of money and data to complete these transactions. In every e-commerce site, consumers can use a recommendation system (RS) to discover stuff like mobile phones, clothing, books, motorcycles, or the recommended item. These technologies are crucial for decision-making because they assist users in maximizing gains or reducing risks. Several modern computer science organizations, like Google, Twitter, LinkedIn, and Netflix, incorporate RSs.

This paper's goal is to make online shopping simpler so that customers can purchase things more simply and sellers can manage their sales more effectively. As previously noted, ML algorithms are being applied in RSs to deliver better suggestions to consumers. The fact that there are so many techniques and modifications that researchers suggest in the literature, however, prevent the ML field from having a clear classification of its algorithms [20]. As a result, selecting an ML method that meets one's needs when designing an RS is challenging and complex. In order to create a better recommendation system, we must identify the best ML algorithm. The use of the GNB, DT, RF, and LR for creating recommendation systems with and without feature reduction using the principal component analysis method is covered in this study (PCA).

## II. LITERATURE REVIEW

The authors of reference [1] proposed how to choose a ML algorithm for product recommendation by collecting publications and analysing them and authors proposed that decision tree and Bayesian algorithms are usually applied in recommendation system.

The authors of reference [2] proposed how to promoting the product on social media for increasing the business performance by K-means machine learning algorithm for product recommendation but this system is dependent on social media platform.

The authors of reference [3] proposed How to help internet retailers increase their revenue by K-means machine learning algorithm for product, but there is no standard rule for choosing correct KPI.

The authors of reference [4] proposed How to transform the amount of blog articles and SSL certificate into search engine traffic for product recommendation for increasing the business performance by Fuzzy-set Qualitative Comparative Analysis.

Research [5] "Predicted performance," This system's analysis aims to enhance performance utilizing fuzzy association rules and better anticipate sales. To estimate sales by type of group for this investigation, data were taken from an online store. Data is classified using modified clustering techniques with fuzzy association rule mining approach for retail based on variables and associated equations implementation. When overlapping and has many clusters for a new item, grouping is done on one object and put in one cluster using the fuzzy approach. The matrix establishes the proximity's size.

An Apriori-Based Method to Product Placement in Order, by Yusuke Ito and Shohei Kato (2016), has been used in a number of information systems studies. In this study, it is stated that the method used is very successful and very effective, but this research was only conducted on small-scale warehouses, not on a large scale [6]. The goal of picking is to make and manage warehouse goods as easily as possible, with the intention of shortening the time in the collection of goods in the warehouse.

## III. METHODOLOGY

The goal of this research on an e-commerce product recommendation system is to provide customers with ideas using machine learning techniques. We have created a dataset from our e-commerce site using customer behaviour this dataset contains user interaction for a certain product. After that we pre-process and filter this dataset. Then applied four machine learning algorithms (GNB, DT, RF, and LR) on this dataset to build a recommendation system model.

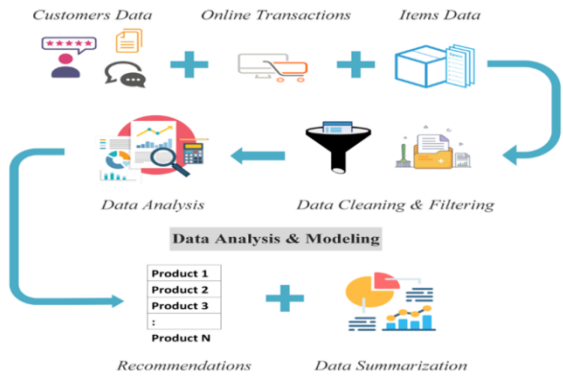

*Fig. 1: The architecture of recommender system*

### A. Dataset Description

This dataset includes the results gathered from a customer survey for an e-commerce platform that was done using a Google Form. The purpose of the survey was to learn more about consumers' preferences, experiences, and satisfaction with the platform's goods and services. The dataset offers insightful insights for enhancing the online shopping experience and making data-driven business choices. It was built on customer sentiment and contains data of transactions. This dataset has 11 features like customer id, name, email, product model, product quantity, product price, customer address, phone number, order date, order status and customer feedback message. We are converted the textual data to numeric values using the label encoding approach to prepare for using machine learning algorithms. Here total 75% data are used for training the model and remainder 25% data are used for testing the model.

### B. Logistic Regression

Logistic Regression uses classification to assess whether an input is benign or not. It is a machine learning approach. Other names for the algorithm include logistic regression, log-linear classifier, and maximum-entropy classifier (MaxEnt). Use of the sigmoid function is made. For the data values A=A1, A2, A3…An, equation 4's calculation of the linear equation is performed. Equation 5 illustrates how to use the

$$W^T = \max \sum_{j=1}^{n} (Y_j \times W_j A_j)$$

$$r = Y_j \times W^T A_j$$

$$P(r) = \frac{1}{1 + \exp^{-r}}$$

### C. Gaussian Naive Bayes

Naive Bayes, a probabilistic classification model for machine learning that is simple but effective, is influenced by the Bayes Theorem. The Bayes theorem is a mathematical formula that provides a conditional probability that an event A will occur provided that an event B has already occurred. The following is its mathematical formula: Naive Bayes, a probabilistic classification model for machine

$$P(A|B) = \frac{P(B|A).P(A)}{P(B)}$$

Where

A and B are two events.

P(A|B) is the probability of event A provided event B has already happened.

P(B|A) is the probability of event B provided event A has already happened.

P(A) is the independent probability of A.

P(B) is the independent probability of B.

### D. Random Forest

In machine learning, RF is a supervise learning algorithm. Different decision trees are trained in this model using the dataset. Because there are many different decision trees involved in this model's decision-making process, it is known as an ensemble of decision trees.

### E. Decision Tree

The decision tree is a well-known machine learning algorithm. It is utilized for data classification. It is a tree-structured algorithm in which internal nodes and branches that indicate decision rules specify the characteristics of a database, with each leaf node expressing the result. The procedures listed below are used to generate a decision tree.

- Choose the target attribute.
- Calculate Information Gain (I.G) for the target attribute
- Calculate the Entropy of the other attributes using the following formula:

$$Entropy(s) = \sum_{i=1}^{n} -(P_i \log_2 P_i)$$

- Subtract Entropy(s) from Information Gain (I.G) of each attribute for find out the Gain (G)

$$Gain(S, A) = Entropy(s) - \sum_{i=1}^{n} \frac{S_v}{S} \times Entropy(S_v)$$

### F. Principal Component Analysis

Algorithm for unsupervised machine learning using principal components (PCA). It is employed to lessen a dataset's dimensionality. It is a statistical method that uses orthogonal transformation to turn observations of correlated features into a collection of linearly uncorrelated data. The Principal Components are the recently modified features.

### G. Performance parameters

We measured the performance of each model by calculating Accuracy, Mean Absolute Error (MAE), Mean Squared Error (MSE), Root Mean Squared Error (RMSE), and R-Squared.

**Accuracy:** One parameter for assessing classification models is accuracy. The percentage of predictions that our model correctly predicted is known as accuracy. The following is the official definition of accuracy:

$$Accuracy = \left(\frac{Number\ of\ correct\ predictions}{Total\ number\ of\ predictions}\right)$$

**R Square/Adjusted R Square:** R Square measures how much of the variation in the dependent variable the model can account for. Its name, R Square, refers to the fact that it is the square of the correlation coefficient (R).

$$R^2 = 1 - \frac{SS_{Regression}}{SS_{Total}} = 1 - \frac{\sum_i (y_i - \hat{y}_i)^2}{\sum_i (y_i - \bar{y})^2}$$

R Square is calculated by dividing the entire sum of the squares that replace the calculated forecast with the mean by the squared prediction error. A higher R Square value denotes a better fit between the prediction and the actual value, which ranges from 0 to 1. To evaluate how well the model fits the dependent variables, use R Square.

**Mean Square Error (MSE):** Mean Square Error is an absolute measurement of the goodness of the fit, whereas R Square is a relative indicator of how well the model fits the dependent variables.

$$MSE = \frac{1}{N} \sum_{i=1}^{N} (y_i - \hat{y}_i)^2$$

MSE is computed by adding together the squares of the real output and the anticipated output, and dividing the result by the total number of data points. It provides you with an exact number indicating how many your findings differ from what you projected. Even while you can't draw many conclusions from a single result, it does provide you with a concrete figure to compare to the outcomes of other. For MSE, there is no ideal value. In other words, the lower the value, the better, and 0 denote a perfect model.

**Mean Absolute Error (MAE):** Mean Square Error and Mean Absolute Error are related terms (MSE). However, MAE takes the total of the error's absolute value rather than its squared sum, as in MSE.

$$MAE = \frac{1}{N} \sum_{i=1}^{N} |y_i - \hat{y}_i|$$

## IV. RESULTS AND DISCUSSION

All the experimental obtained results for this work are exhibited in this section in the tabular form and the obtained results are analyzed by performance evaluation parameters. Four machine learning algorithms like RF, DT, GNB and LR can be used for required evaluation. The results are presented in the following figure. In this measurement, the **Random Forest (RF) gives highest accuracy 99.8%,** the Decision Tree (DT) exhibits 96.3%, the Gaussian Naive Bayes (GNB) gives 44.4%, and the Logistic Regression (LR) shows 22.024%, the least accuracy.

Fig.2: Shows the ML model's accuracy performance.

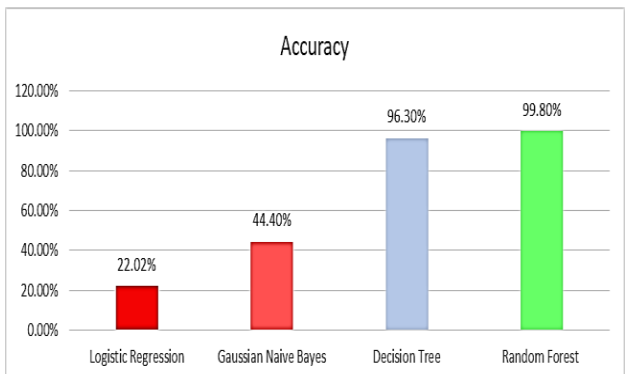

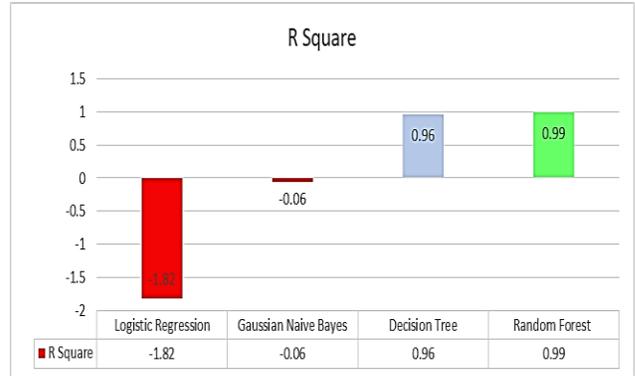

Fig. 3: ML model's R square performance

In this measurement, the Random Forest (RF) achieves the highest value at 0.99, followed by the Decision Tree (DT) at 0.96. The Gaussian Naive Bayes (GNB) records a value of -0.06, and the Logistic Regression (LR) shows -1.82.

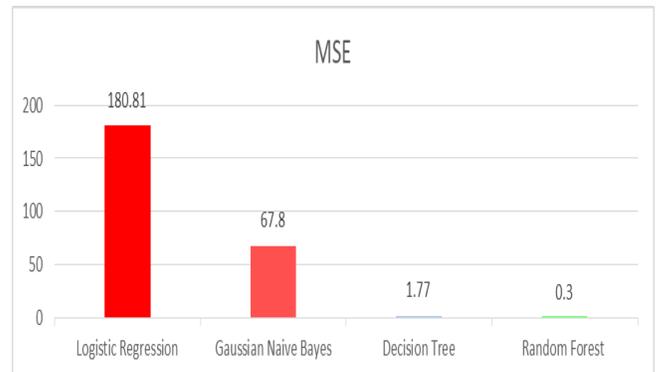

Fig. 4: ML model's MSE value.

In this measurement, the Random Forest (RF) shows the best value at 0.3, followed by the Decision Tree (DT) at 1.77. The Gaussian Naive Bayes (GNB) gives a value of 67.8, while the Logistic Regression (LR) shows 180.81.

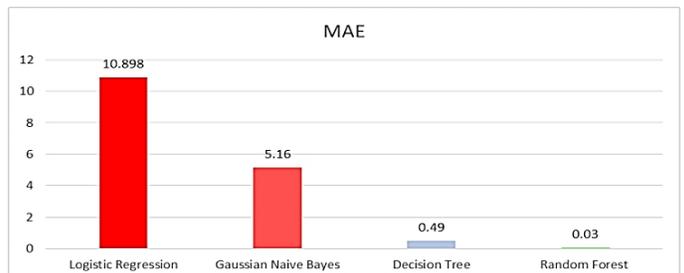

In this measurement, the Random Forest (RF) shows the best value at 0.03, followed by the Decision Tree (DT) at 1.49. The Gaussian Naive Bayes (GNB) records a value of 5.16, and the Logistic Regression (LR) shows 10.898.

Table 1. : Overall effectiveness of the LR.

| Model | Accuracy | R Square | MSE | MAE |
|---|---|---|---|---|
| LR | 24.3% | -1.63 | 169.1 | 10.3 |
|  | 19.12% | -1.93 | 193.7 | 11.44 |
|  | 23.5 | -1.79 | 178.05 | 10.8 |
|  | 21.5% | -2.01 | 191.4 | 11.38 |
|  | 19.12% | -1.9 | 182.2 | 11.08 |
|  | 24.7% | -1.73 | 170.3 | 10.49 |
|  | 21.5% | -1.76 | 179.05 | 10.8 |
|  | 21.5% | -1.766 | 179.88 | 10.85 |
|  | 23.5% | -1.83 | 187.36 | 11.11 |
|  | 21.5% | -1.75 | 177.1 | 10.73 |
|  | 22.024% | -1.82 | 180.81 | 10.898 |

Table 2. : Overall effectiveness of the GNB.

| Model | Accuracy | R Square | MSE | MAE |
|---|---|---|---|---|
| GNB | 40.23% | -0.03 | 66.55 | 5.26 |
|  | 44.6% | 0.28 | 47.3 | 3.94 |
|  | 43.4% | -0.59 | 101.7 | 6.82 |
|  | 41.03% | 0.02 | 61.65 | 4.94 |
|  | 60% | 0.35 | 39.8 | 3.4 |
|  | 38.2% | -0.60 | 99.6 | 6.9 |
|  | 42.2% | -0.42 | 92.30 | 6.13 |
|  | 49.4% | 0.359 | 41.6 | 3.58 |
|  | 41.8% | 0.132 | 57.27 | 5.01 |
|  | 40.6% | -0.098 | 70.66 | 5.581 |
|  | 44.4% | -0.06 | 67.8 | 5.16 |

Table 3. : Overall effectiveness of the DT.

| Model | Accuracy | R Square | MSE | MAE |
|---|---|---|---|---|
| DT | 96.6% | 0.96 | 0.14 | 2.0 |
|  | 97.2% | 0.96 | 0.14 | 2.0 |
|  | 97.5% | 0.97 | 0.016 | 0.11 |
|  | 99.8% | 0.99 | 0.063 | 0.015 |
|  | 99.8% | 0.99 | 0.063 | 0.015 |
|  | 87.1% | 0.87 | 7.98 | 0.34 |
|  | 100% | 1 | 0.0 | 0.0 |
|  | 97.6% | 0.976 | 1.537 | 0.09 |
|  | 96.1% | 0.96 | 2.55 | 0.115 |
|  | 92.9% | 0.929 | 4.625 | 0.227 |
|  | 96.3% | 0.96 | 1.77 | 0.49 |

Table 4. : Overall effectiveness of the RF.

| Model | Accuracy | R Square | MSE | MAE |
|---|---|---|---|---|
| RF | 99.6% | 0.96 | 0.19 | 0.08 |
|  | 100% | 1 | 0.0 | 0.0 |
|  | 100% | 1 | 0.0 | 0.0 |
|  | 100% | 1 | 0.0 | 0.0 |
|  | 99.2 | 0.96 | 0.41 | 0.13 |
|  | 100% | 1 | 0.0 | 0.0 |
|  | 99.6% | 0.992 | 0.48 | 0.04 |
|  | 100% | 1 | 0.0 | 0.0 |
|  | 99.6% | 0.97 | 1.92 | 0.087 |
|  | 100% | 1 | 0.0 | 0.0 |
|  | 99.8% | 0.988 | 0.3 | 0.03 |

The Random Forest (RF) model yields the best results in this observation. In contrast, the Decision Tree (DT) shows a 0.96 R-square value, 1.77 MSE, and 0.49 MAE, with a 99.3% accuracy. Logistic Regression (LR) demonstrates the lowest accuracy at 22.02%, with a -1.82 R-square value, 180.81 MSE, and 10.898 MAE. Gaussian Naive Bayes (GNB) has 44.4% accuracy, a -0.06 R-square value, 67.8 MSE, and 5.16 MAE. The Random Forest (RF) exhibits the highest performance with 99.8% accuracy, a 0.99 R-square value, 0.3 MSE, and 0.03 MAE. Thus, among all the techniques, the

Random Forest (RF) algorithm is the most effective in predicting the outcome.

## V. CONCLUSION

Nowadays, recommender systems (RS) are widely utilized in social networks, e-commerce, and various other fields. The integration of machine learning (ML) methods, which enable computers to learn from user input and provide more personalized suggestions, marks a significant advancement in the development of RS. This research explores four machine learning approaches for recommendation engines: Decision Tree (DT), Gaussian Naive Bayes (GNB), Random Forest (RF), and Logistic Regression (LR).To optimize outcomes and save time, the study employs the Principal Component Analysis (PCA) method for feature reduction. The performance of each model is evaluated using multiple assessment metrics, including accuracy, R-square score, Mean Squared Error (MSE), and Mean Absolute Error (MAE). Experimental results indicate that the RF algorithm outperforms the GNB, DT, and LR algorithms, achieving the highest accuracy.

**Data Availability**
**Data generated for this study is available from the corresponding author on formal request.**